\documentclass[aps, pra, twocolumn,showpacs,preprintnumbers,amsmath,amssymb]{revtex4}
\newcommand{\ket}[1]{\mbox{$ | #1 \rangle $}}

\begin{document}

\preprint{}

\title{Quantum key distribution protocols using entangled state}
% Force line breaks with \\

\author{Jian Wang}

 \email{jwang@nudt.edu.cn}

\affiliation{School of Electronic Science and Engineering,
\\National University of Defense Technology, Changsha, 410073, China }
\author{Quan Zhang}
\affiliation{School of Electronic Science and Engineering,
\\National University of Defense Technology, Changsha, 410073, China }
\author{Chao-jing Tang}
\affiliation{School of Electronic Science and Engineering,
\\National University of Defense Technology, Changsha, 410073, China }


\begin{abstract}
We present three quantum key distribution protocols using entangled
state. In the first two protocols, all Einstein-Podolsky-Rosen pairs
are used to distribute a secret key except those chosen for
eavesdropping check, because the communication parties measure each
of their particles in an invariable measuring basis. The first
protocol is based on the ideal of qubit transmission in blocks.
Although it need quantum memory, its theoretic efficiency
approximates to 100\%. The second protocol does not need quantum
memory and its efficiency for qubits can achieve 100\%. In the third
protocol, we present a controlled quantum key distribution using
three-particle entangled state to solve a special cryptographic
task. Only with the controller's permission could the communication
parties establish their sharing key and the sharing key is secret to
the controller. We also analyze the security and the efficiency of
the present protocols.
\end{abstract}

%Use showkeys class option if keyword
\maketitle

Quantum key distribution (QKD) is one of the most promising
applications of quantum information science. The goal of QKD is to
allow two legitimate parties, Alice and Bob, to generate a secret
key over a long distance, in the presence of an eavesdropper, Eve,
who interferes with the signals. The security of QKD is based on the
fundamental laws of physics. Together with the Vernam cipher, QKD
can be used for unconditionally secure communication. Since the BB84
protocol\cite{bb84}, the first QKD scheme, was published, many
variations on QKD have been subsequently proposed. They can be
roughly classified into ``prepare and measure'' protocols, such as
BB84, B92\cite{b92}, the three-state protocol\cite{ha00}, the
six-state protocol\cite{b98} and ``entanglement based'' protocols,
such as E91\cite{e91}, BBM92\cite{bbm92}. There have been efforts to
set a security proof based on entanglement for the both
classes\cite{proof}. Recently, the continuous-variable
QKD\cite{continues} has also been proved to be a promising protocol
to send secret keys with high transmission rate.

The efficiency is one of the important parameters of QKD protocol.
Many efforts have been made to improve the efficiency of QKD
protocol. To improve the efficiency of the BB84 QKD, the scheme in
Ref.\cite{hhm} assigns significantly different probabilities for the
different polarization bases during both transmission and reception
to reduce the fraction of discarded data. The actual probabilities
used in their schemes are announced in public. To defeat the
eavesdropper's attack to the predominant basis, it needs a refined
analysis of accepted data: they separate the accepted data into
various subsets according to the basis employed and estimate an
error rate for each subset individually. In Ref.\cite{wiy98}, Hwang,
Koh, and Han proposed a modified BB84 QKD scheme that increases its
efficiency to nearly 100\%. However, the communication parties need
a common secret key in their scheme. From the point of view of
information theory, Cabello\cite{cabello1} defined the efficiency of
a QKD protocol, $\mathcal{E}$,
\begin{eqnarray}
\mathcal{E}=\frac{b_s}{q_t + b_t},
\end{eqnarray}
where $b_s$ is the number of secret bits received by Bob, $q_t$,
$b_t$ is each the number of qubits, classical bits interchanged
between Alice and Bob during the QKD process. Here the classical
bits used for eavesdrop checking have been neglected. As has been
discussed by Cabello, the efficiency of BB84, E91, cabello
2000\cite{cabello2}, is 25\%, 50\%, 67\%, respectively. Actually,
qubit is more expensive than classical bit. The efficiency equation
$\mathcal{E}$ (called the total efficiency) cannot describe the
efficiency of QKD protocol sufficiently. The efficiency for qubits
is a useful complement to analyze the efficiency of QKD protocol,
which is defined as
\begin{eqnarray}
\eta=\frac{q_u}{q_t},
\end{eqnarray}
where $q_u$ is the useful qubits and $q_t$ is the total qubits
transmitted \cite{deng}. To evaluate the efficiency of a QKD
protocol, we should combine these two parameters.

In this paper, we present three QKD protocols using entangled state.
The first protocol whose ideal is based on qubit transmission in
blocks uses Einstein-Podolsky-Rosen (EPR) pairs to distribute a
secret key. The theoretic efficiency of the protocol approximates to
100\%, because all EPR pairs are used to distribute a secret key
except those chosen for checking eavesdroppers. However, the flaw of
the first protocol is that it needs quantum memory during the QKD
process. We then present the second protocol which does not need
quantum memory. In the second protocol, although the total
efficiency is 50\%, the efficiency for qubits approaches 100\%. To
solve a special cryptographic task, we propose a controlled QKD
protocol. The two communication parties can only generate their
sharing key with the permission of the controller. This protocol can
also be used to distribute a secret key among three parties. The
three protocols are unconditionally secure.

%%%%%%%%%%%%%%%%%%%%%%%%%%%%%%%%%%%%%%%%%%%%%%%%%%%%%%%%%%%%%%%%%%%%%
%
G. L. Long and X. S. Liu proposed an efficient high-capacity QKD
scheme\cite{ll02} whose efficiency can achieve 100\%. In their
scheme, a set of ordered $N$ EPR pairs is used as a data block for
distributing secret key. Based on the ideal of block transmission,
we present another QKD protocol whose efficiency can also achieve
100\%. The security of the present protocol is ensured by the random
Hadamada transformation. Our protocol is as follows:

\textbf{Protocol 1:}

(1) Alice prepares an ordered $N$ EPR pairs in the Bell
state
\begin{eqnarray}
\ket{\phi^+}_{AB}=\frac{1}{\sqrt{2}}(\ket{00}+\ket{11})_{AB}.
\end{eqnarray}
We denotes the ordered $N$ EPR pairs with \{[P$_1(A)$,P$_1(B)$],
[P$_2(A)$,P$_2(B)$], $\cdots$, [P$_N(A)$,P$_N(B)$]\}, where the
subscript indicates the pair order in the sequence, and $A$, $B$
represents the two particles of EPR pair, respectively. Alice takes
one particle from each EPR pair to form an ordered EPR partner
particle sequence [P$_1(A)$, P$_2(A)$,$\cdots$, P$_N(A)$], called
$A$ sequence. The remaining EPR partner particles compose $B$
sequence, [P$_1(B)$, P$_2(B)$,$\cdots$, P$_N(B)$]. Alice transmits
the $B$ sequence to Bob.

(2) To prevent eavesdropping, Bob selects randomly a sufficiently
large subset of $B$ sequence and performs Hadamard transformations
on them. He then announces publicly the position of the selected
particles. The Hadamard transformation is crucial for the security
of the scheme as we will see in the sequel.

(3) After hearing from Bob, Alice executes Hadamard transformations
on the corresponding particles of $A$ sequence.

(4) Note that
\begin{eqnarray}
\ket{\phi^+}_{AB}=\frac{1}{\sqrt{2}}(\ket{++}+\ket{--})_{AB},
\end{eqnarray}
where $\ket{+}=\frac{1}{\sqrt{2}}(\ket{0}+\ket{1})$,
$\ket{-}=\frac{1}{\sqrt{2}}(\ket{0}-\ket{1})$. The two parties'
Hadamada transformations will not change the initial state. Alice
then measures the $A$ sequence in $Z$-basis \{\ket{0}, \ket{1}\}.
Bob measures the $B$ sequence in the same basis as Alice. Thus they
established sharing key. In this step, they can also measure each of
their particles in $X$-basis \{\ket{+}, \ket{-}\}.

(5) Alice and Bob then publicly compare the results of these
measurements to check eavesdropping. Bob chooses randomly a
sufficiently large subset of his results and announces them
publicly. Alice compares Bob's results with her corresponding
results. She can thus find out whether there is an eavesdropper. If
too many of these measurements disagree, they abort the protocol. If
they are certain that there is no eavesdropping, Alice and Bob
utilize privacy amplification and error correction to distil the
final key.

We now discuss the security and the efficiency of protocol 1. The
crucial point is that the Hadamard transformations at the step 2 and
3 of the protocol do not allow Eve to have a successful attack and
Eve's attack will be detected during the eavesdropping check. The
protocol 1 is similar to the modified Lo-Chau protocol \cite{shor},
but the protocol 1 does not need classical message except that used
for eavesdropping check. The security of the protocol 1 is the same
as the modified Lo-Chau protocol which is proved unconditionally
secure. According to the information-theoretic efficiency defined by
Cabello, the total efficiency of protocol 1 can be made
asymptotically close to 100\%. Here the classical bits used for
eavesdrop checking have been neglected. The efficiency for qubit can
also achieve 100\% because all EPR pairs are used to distribute a
sharing key except those used to check eavesdropping.

Although the protocol 1 is efficient, it is necessary for the
protocol to use quantum memory. We then present another QKD protocol
without quantum memory.

\textbf{Protocol 2:}

(1) Alice prepares randomly an EPR pair in one of the following
states
\begin{eqnarray}
\ket{\phi^\pm}_{AB}=\frac{1}{\sqrt{2}}(\ket{00}\pm\ket{11})_{AB},
\end{eqnarray}
where the subscripts $A$, $B$ represents the two particles of EPR
pairs. Alice then send the particle $B$ to Bob.

(2) After hearing from Bob, Alice publishes the information of the
initial state she prepared. If the initial state is \ket{\phi^-},
both Alice and Bob perform Hadamard transformation on each of their
corresponding particle. Otherwise, they do nothing. After Hadamada
transformation, \ket{\phi^-} is changed to
\begin{eqnarray}
\ket{\phi^-_1}_{AB}=\frac{1}{\sqrt{2}}(\ket{++}-\ket{--})_{AB},
\end{eqnarray}

(3) After doing this, Alice and Bob measure their corresponding
particles in $X$-basis. They agree that \ket{+} (\ket{-})
corresponds to bit ``0'' (``1'').

(4) The parties repeat the above steps $N$ times and generate $N$
raw secret keys.

(5) To check eavesdropping, Alice selects randomly a sufficiently
large subset of her measurement results and tells it to Bob. Bob
publishes his measurement results of the sampling particles. Alice
then evaluates the error rate of the QKD process. If the error rate
exceeds the threshold, they abort the protocol. Otherwise, they
utilize privacy amplification and error correction to distil a final
key.

We then analyze the security and the efficiency of protocol 2.
Firstly, the protocol is secure against the intercept-resend attack
by Eve. In this attack, Eve intercepts the particle $B$ and makes
measurement on it, then she resends a particle to Bob according to
her measurement result. Eve can only intercept the particle $B$ at
the step 1 of the protocol and she cannot make certain which
particle will be executed Hadamard transformation. Thus Eve can only
measure the intercepted particle in $Z$-basis or $X$-basis randomly.
Suppose Eve measures the intercepted particle which belongs to
\ket{\phi^+} in $Z$-basis. If the result of Eve's measurement is
``0'', she sends a particle in the state $\ket{+}$ to Bob, otherwise
sends a particle in the state $\ket{-}$. Then the state of the two
particles collapses to $\ket{0+}_{AB}$ or $\ket{1-}_{AB}$ each with
probability 1/2. Thus the error rate introduced by Eve will achieve
50\%. During the eavesdropping check, Eve's attack will be detected.
Suppose Eve performs $X$-basis measurement on the intercepted
particle which belongs to \ket{\phi^-}. Then the state of the two
particles collapses to $\ket{+-}_{AB}$ or $\ket{-+}_{AB}$ each with
probability 1/2. After Hadamada transformation, the state is changed
to $\ket{01}_{AB}$ or $\ket{10}_{AB}$. The error rate introduced by
Eve will also achieve 50\%.

Secondly, the protocol is safe against collective attack. In this
strategy, Eve intercepts the particle $B$ and uses it and her own
ancillary particle in the state \ket{0} to do a CNOT operation (the
particle $B$ is the controller, Eve's ancillary particle is the
target). Then Eve resends the particle $B$ to Bob. However, Eve has
no information of the initial state. Suppose the initial state is
\begin{eqnarray}
\ket{\phi^-}_{AB}=\frac{1}{\sqrt{2}}(\ket{+-}+\ket{-+})_{AB}.
\end{eqnarray}
After Eve's collective attack, the state of the particle $A$, $B$
and Eve's ancillary particle becomes
\begin{eqnarray}
\ket{\Omega_1}_{ABE}=\frac{1}{\sqrt{2}}(\ket{+-1}+\ket{-+0})_{ABE},
\end{eqnarray}
where the subscript $E$ indicates Eve's ancillary particle.
According to the protocol, Alice and Bob perform Hadamada
transformations on their corresponding particles, obtaining
\begin{eqnarray}
\ket{\Omega_2}_{ABE}=\frac{1}{\sqrt{2}}(\ket{011}+\ket{100})_{ABE}.
\end{eqnarray}
Obviously, Eve's eavesdropping will be detected during the
eavesdropping check because half of Bob's results will be
inconsistent with that of Alice's. As we described above, the total
error rate introduced by Eve is 25\%. Actually, the security of the
protocol 2 is equal to that of the modified Lo-Chau protocol. We can
also give a Shor-Preskill-type proof to protocol 2.

In the protocol 2, the efficiency for qubits is 100\% because all
EPR pairs are used to generate a key except those chosen for
eavesdropping check. It is not necessary for the protocol 2 to use
quantum memory. However, it needs a bit of classical message to
generate a bit of sharing key. Therefor the total efficiency of the
protocol 2 is 50\%. Certainly, we should pay more attention to the
efficiency for qubits because qubit is more expensive than classical
bit.

Han et al. proposed a controlled QKD scheme with three-particle
entanglement \cite{han}. According to their scheme, the Eq. 1 of
their scheme should be
\begin{eqnarray}
\ket{ABC}&=&\frac{1}{2}(\ket{HHH}+\ket{HVV}+\ket{VHV}-\ket{VVH})\nonumber\\
&=&\frac{1}{2}[\ket{H}(\ket{HH}+\ket{VV})\nonumber\\
& &+\ket{V}(\ket{HV}-\ket{VH})]
\end{eqnarray}
Suppose Bob intercepts the two photons which Alice sends to Bob and
Carol. He then performs Bell basis measurement on the intercepted
photons and obtains $\ket{\phi^+}$ and $\ket{\psi^-}$ each with
probability 1/2. Bob will obtain Alice's measurement result and his
action will not be detected by Alice. Bob then resends one of the
intercepted particles to Carol. According to their scheme, Bob and
Carol can establish sharing key without the control of Alice. To
solve this problem, we propose another controlled QKD protocol.
Suppose only Alice can prepare the entangled state. Bob and Charlie
can only generate their sharing key under the control of the
controller Alice. Without Alice's permission, Bob and Charlie cannot
establish their sharing key. Certainly, the sharing key is secret to
Alice.

\textbf{Protocol 3:}

(1) Alice prepares a three-particle entangled state in the state
\begin{eqnarray}
\ket{\Psi_1}_{ABC}=\frac{1}{\sqrt{2}}(\ket{0+0}+\ket{1-1})_{ABC}
\end{eqnarray}
or
\begin{eqnarray}
\ket{\Psi_2}_{ABC}=\frac{1}{\sqrt{2}}(\ket{0-0}+\ket{1+1})_{ABC}
\end{eqnarray}
randomly. She then sends the particle $B$, $C$ to Bob and Charlie,
respectively.

(2) After confirming that Bob and Charlie have received their
particles, Alice measures the particle $A$ in $Z$-basis or $X$-basis
randomly. She then publishes his measuring basis. If Alice performed
$Z$-basis measurement, Bob measures the particle $B$ in the
$X$-basis and Charlie performs $Z$-basis measurement on the particle
$C$. If Alice performs $X$-basis measurement, Bob (Charlie) performs
$Z$-basis ($X$-basis) measurement on his particle.

(3) The parties repeat the above steps $N$ times and each obtains
$N$ measurement results. The value of $N$ is large enough to check
eavesdropping, but not enough to generate a key which is long enough
to ensure the security of cryptography. To generate a secure key,
the parties should run the protocol twice at least.

(4) Alice chooses randomly a sufficiently large subset of her
measurement results and announces the positions of the sampling
particles publicly. She let Bob and Charlie publish their
measurement results of the sampling particles. Note that
\begin{eqnarray}
\label{1}
\ket{\Psi_1}_{ABC}&=&\frac{1}{\sqrt{2}}[\ket{+}(\ket{0+}+\ket{1-})\nonumber\\
& &+\ket{-}(\ket{0-}+\ket{1+})]_{ABC}
\end{eqnarray}
and
\begin{eqnarray}
\label{2}
\ket{\Psi_2}_{ABC}&=&\frac{1}{\sqrt{2}}[\ket{+}(\ket{0+}-\ket{1-})\nonumber\\
& &+\ket{-}(\ket{0-}-\ket{1+})]_{ABC}.
\end{eqnarray}
They can make certain whether there exists eavesdropping by
comparing their measurement results. If Bob performs Bell basis
measurement on the particle $B$ and $C$, his action will also be
detected by Alice. If there is no eavesdropping, they continue to
the next step. Otherwise, the protocol is halted.

(5) If Alice permits Bob and Charlie to establish their sharing key,
she publishes her measurement results of the particles on which she
performed $X$-basis measurements. The parties let $\ket{0}$,
$\ket{+}$ correspond to binary ``0'' and $\ket{1}$, $\ket{-}$
correspond to binary ``1''. If Alice's measurement result is
\ket{+}, Bob and Charlie obtains a identical raw key ``0'' or ``1''.
If Alice's measurement result is \ket{-}, Bob or Charlie should
invert the bit value of the key to obtain a identical key.
Obviously, the key is secret to the controller Alice. If Alice, Bob
and Charlie want to establish a three-party key, it only needs Alice
to publish the initial state of the particles on which Alice
performed $Z$-basis measurement. If the initial state is
\ket{\Psi_2}, Bob should invert his bit of the key. They can also
distil a final key using privacy amplification and error correction.

We now discuss the security of the protocol 3. The two state
\ket{\Psi_1}, \ket{\Psi_2} which are prepared randomly by Alice,
ensure that Bob and Charlie cannot establish sharing key without
Alice's permission. Suppose Bob intercepts the particle $B$ and $C$
and performs Bell basis measurement on the intercepted particles.
Note that
\begin{eqnarray}
\ket{0+}+\ket{1-}=\frac{1}{\sqrt{2}}(\ket{00}-\ket{11}+\ket{01}+\ket{10})\nonumber\\
\ket{0-}+\ket{1+}=\frac{1}{\sqrt{2}}(\ket{00}+\ket{11}-\ket{01}+\ket{10})\nonumber\\
\ket{0+}-\ket{1-}=\frac{1}{\sqrt{2}}(\ket{00}+\ket{11}+\ket{01}-\ket{10})\nonumber\\
\ket{0-}-\ket{1+}=\frac{1}{\sqrt{2}}(\ket{00}-\ket{11}-\ket{01}-\ket{10}).
\end{eqnarray}
According to the Eq. \ref{1} and \ref{2}, Bob cannot make certain
the state of the particle $B$, $C$. He also has no information of
Alice's result. During the eavesdropping check, his action will be
detected by Alice. Alice then stops the protocol. Because the
parties have run the protocol only once, Bob and Charlie cannot
generate a secure sharing key even if they obtained some random EPR
pairs. Without regard to the special cryptographic task, the
security of the present protocol can be reduced to that of the BBM92
protocol because the parties measure their corresponding particles
in $Z$-basis or $X$-basis randomly.

To improve the efficiency of the protocol 3, we can use the same
method of the Ref. \cite{han}, which Alice measures her particle in
the $Z$-basis ($X$-basis) with probabilities $\varepsilon$
($1-\varepsilon$), where $0<\varepsilon\leq1$. Obviously, it needs a
refined data analysis to ensure the security of the protocol.

In summary, we propose three QKD protocols with entangled state. It
is not necessary for the protocol 1 and protocol 2 to use
alternative measuring basis. Thus all EPR pairs of the two protocols
are used to generate a secret key except those chosen for
eavesdropping. By using block transmission, the total efficiency of
the protocol 1 can achieve 100\% but it needs quantum memory. The
protocol 2 does not need quantum memory and the efficiency for
qubits is 100\% because the parties measure their particles in an
invariable measuring basis. But the total efficiency of the protocol
2 is 50\% for the use of classical bit. In view of the fact that
qubit is more expensive than classical bit, more attention should be
paid to the efficiency for qubits. The protocol 3 is a controlled
QKD protocol which can be applied to a special cryptographic task.
Bob and Charlie establish their sharing key under the control of
Alice. Only with Alice's permission could Bob and Charlie generate
an identical key. The protocol 3 is multifunctional, which can also
be used to distribute a key among three parties. It is appropriate
for QKD network.

%%%%%%%%%%%%%%%%%%%%%%%%%%%%%%%%%%%%%%%%%%%%%%%%%%%%%%%%%%%%%%%%%%%%%%

This work is supported by the National Natural Science Foundation of
China under Grant No. 60472032.

%%%%%%%%%%%%%%%%%%%%%%%%%%%%%%%%%%%%%%%%%%%%%%%%%%%%%%%%%%%%%%%%%%%%%
%
%

%%%%%%%%%%%%%%%%%%%%%%%%%%%%%%%%%%%%%%%%%%%%%%%%%%%%%%%%%%%%%%%%%%%%%
%
%
\end{document}